\documentclass[rnote]{aa} 
\usepackage{graphicx}
\usepackage{txfonts}
\usepackage{natbib}
\bibpunct{(}{)}{;}{a}{}{,}
\begin{document}
   \title{Resolving the L/T transition binary SDSS J2052-1609\,AB
\thanks{Based on observations collected at the European Southern Observatory, Paranal, Chile, under program 083.C-0305(A). This work is partly based on observations made with the NASA/ESA Hubble Space Telescope, obtained at the Space Telescope Science Institute (STScI) and is associated with program GO-11136. STScI is operated by the Association of Universities for Research in Astronomy, Inc., under NASA contract NAS 5-26555.}}

	\author{M.B. Stumpf\inst{1}\and
		K. Gei{\ss}ler\inst{2} \and
		H. Bouy\inst{3} \and
                W. Brandner\inst{1} \and 
		B. Goldman\inst{1}  \and
		  Th. Henning\inst{1} }
		  

	\institute{
$^{1}$ Max-Planck-Institut f\"ur Astronomie, K\"onigstuhl 17, D-69117 Heidelberg, Germany\\
              \email{stumpf@mpia.de, geissler@astro.sunysb.edu}\\
$^{2}$ Department of Physics and Astronomy, State University of New York, Stony Brook, NY 11794-3800, USA\\
$^{3}$ Centro de Astrobiolog\'\i a, INTA-CSIC, PO BOX 78, E-28691, Villanueva 
de la Ca\~nada, Madrid, Spain\\}

	\date{Received 15 July 2010 / Accepted  25 September 2010 }

\abstract{Binaries provide empirical key constraints for star formation theories, like the overall binary fraction, mass ratio distribution and the separation distribution. They play a crucial role to calibrate the output of theoretical models, like absolute magnitudes, colors and effective temperature depending on mass, metallicity and age.}
{We present first results of our on-going high-resolution imaging survey of late type brown dwarfs. The survey aims at resolving tight brown dwarf binary systems to better constrain the T dwarf binary fraction. We intent to follow-up the individual binaries to determine orbital parameters.}
{Using NACO at the VLT we performed AO-assisted near-infrared observations of SDSS J2052-1609. High-spatial resolution images of the T1 dwarf were obtained in \emph{H} and \emph{K$_\mathrm{S}$} filters.}
{We resolved SDSS J2052-1609 into a binary system with a separation of 0.1009\,$\pm$\,0.001\arcsec. Archival data from HST/NICMOS taken one year previous to our observations proves the components to be co-moving. Using the flux ratio between the components we infer \emph{J}, \emph{H} and \emph{K$_\mathrm{S}$} magnitudes for the resolved system. From the near-IR colors we estimate spectral types of T1\,$^{+\,1}_{-\,4}$ and T2.5\,$\pm$\,1 for component A and B, respectively. A first estimate of the total system mass yields M$_{tot}$\,$\geq$\,78\,M$_{Jup}$, assuming a circular orbit.}
	 {}


\maketitle

\section{Introduction}
Observational studies of binaries provide the key to assess fundamental astrophysical 
parameters like mass, radius or density of compact objects for a wide range of astronomical 
objects ranging from asteroids to brown dwarfs and stars. For brown dwarfs, the continuous 
cooling with increasing age results in a degeneracy of the mass-luminosity relation, hence 
for a brown dwarf of unknown age, dynamical mass estimates based on the determination of 
orbital parameters of a binary provide the only precise means to derive masses.

After the discovery of the first binary brown dwarfs (\citealp{Mart99_1, Mart99_2}), systematic surveys led to the discovery of more than 70 brown dwarfs in binary or multiple systems (e.g., \citealp{bouy2003, burgasser2003b, gizis2003}). Astrometric follow-up observations of some of the shorter period binaries provided first system mass estimates (e.g. \citealp{bouy2004, konopacky2010}), and revealed that mass-luminosity predictions based on theoretical evolutionary models for brown dwarfs still needed to be calibrated against observations. 

Binaries have also been invoked to explain spectral and flux peculiarities observed in brown dwarfs at the transition between L and T dwarfs. \cite{burgasser2006c} find a higher incident of binaries among L/T transition objects compared to dwarfs of earlier or later spectral type, and suggest that spectral features of several L/T transition brown dwarfs could be explained by a superposition of spectra of a mid to late L and a T dwarf. Meanwhile, \cite{goldman2008} concluded that the larger binary fraction among L/T brown dwarfs is not statistically significant and that a much larger observation sample is required.\\
\cite{Liu06} reported the first detection of a flux ratio reversal between primary and secondary in a T dwarf binary in the \emph{J}-\,band. Several more flux reversal L/T transition binaries have been identified in the meantime, supporting the idea of this so-called ``crypto-binarity" (see e.g., \citealp{looper2008}, and references therein; \citealp{stumpf2010}). Hence, binaries composed of a late L dwarf and an early T dwarf could nicely explain both the observed near infrared spectral energy distribution and the spectral characteristics of L/T transition dwarfs.
\begin{figure*}[t!]
\centering
\setlength{\unitlength}{1cm}
\hspace{-1.75cm}
   \begin{minipage}[t]{4.5cm} 
       \includegraphics[scale=0.28]{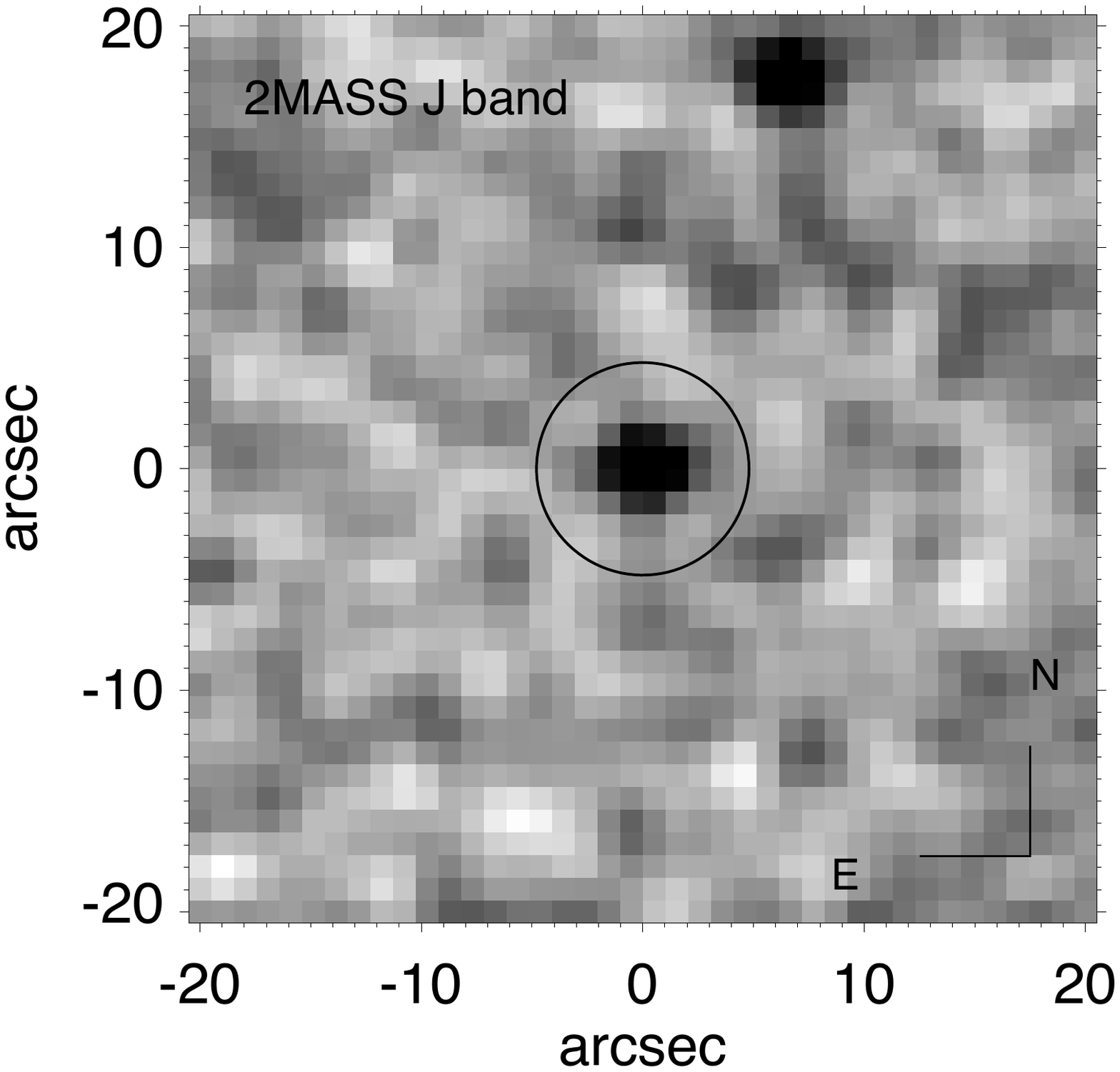}\vspace{-0.6cm}
       \hspace*{2.3cm}   01\,/\,06\,/\,1999
    \end{minipage}\hspace{0.45cm}
    \begin{minipage}[t]{4.5cm} 
        \includegraphics[scale=0.28]{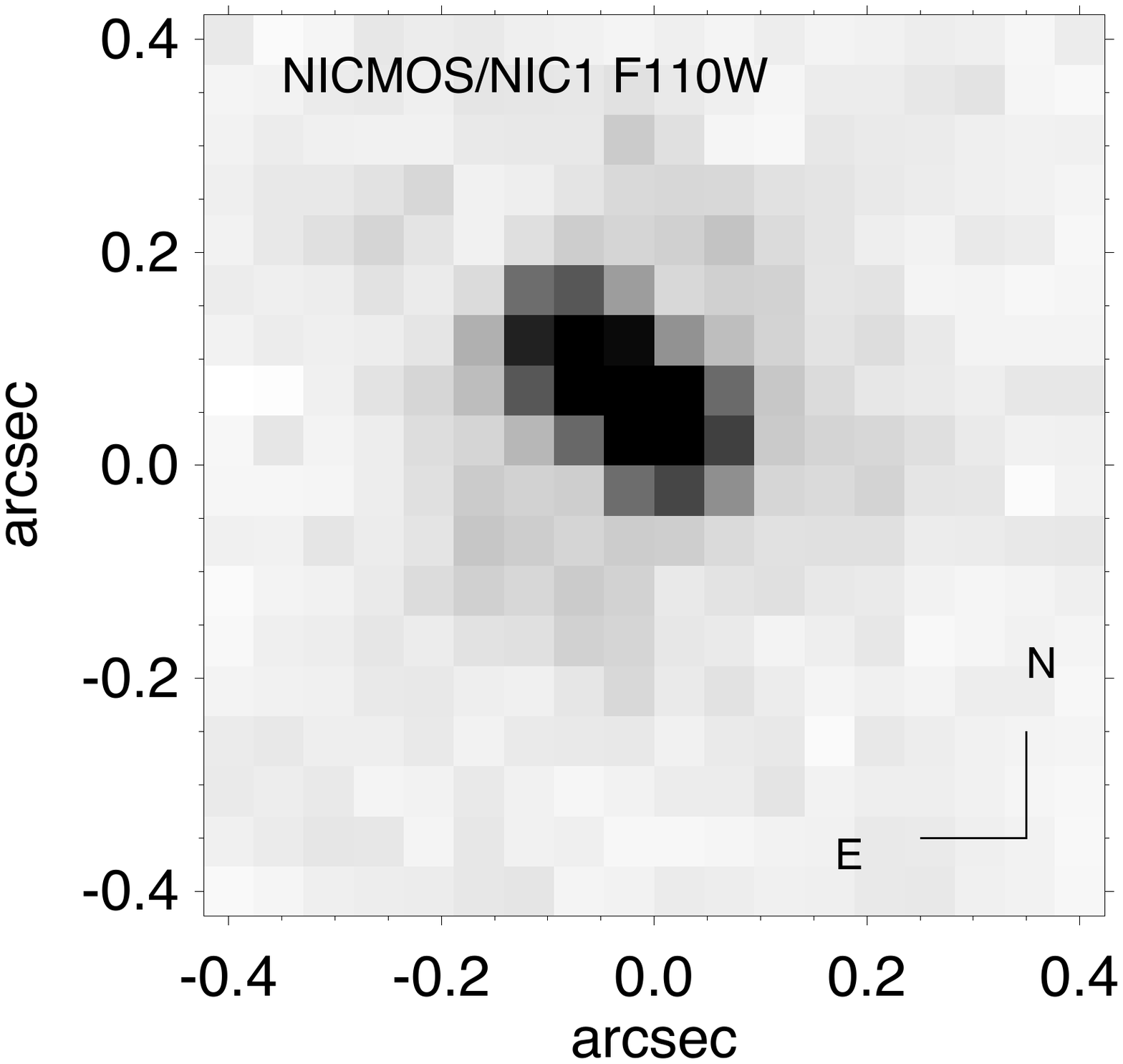}\vspace{-0.6cm}
        \hspace*{2.3cm}  24\,/\,06\,/\,2008
    \end{minipage}\hspace{0.45cm}
    \begin{minipage}[t]{4.5cm} 
        \includegraphics[scale=0.28]{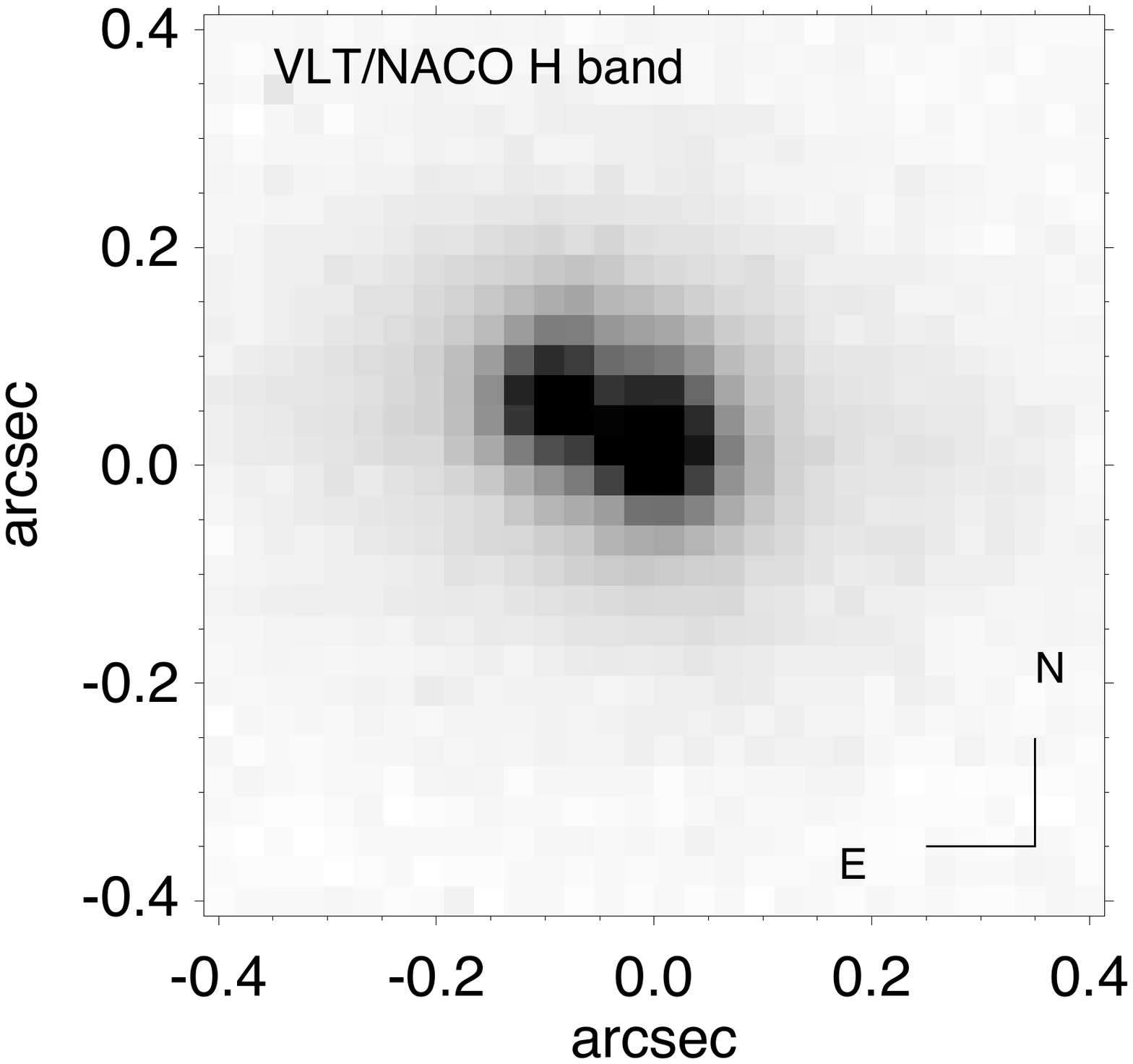}\vspace{-0.6cm}
        \hspace*{2.3cm}  19\,/\,06\,/\,2009
   \end{minipage}\hspace{0.45cm}
    \begin{minipage}[t]{3.5cm} 
    \vspace{-5.35cm}
        \includegraphics[scale=0.255]{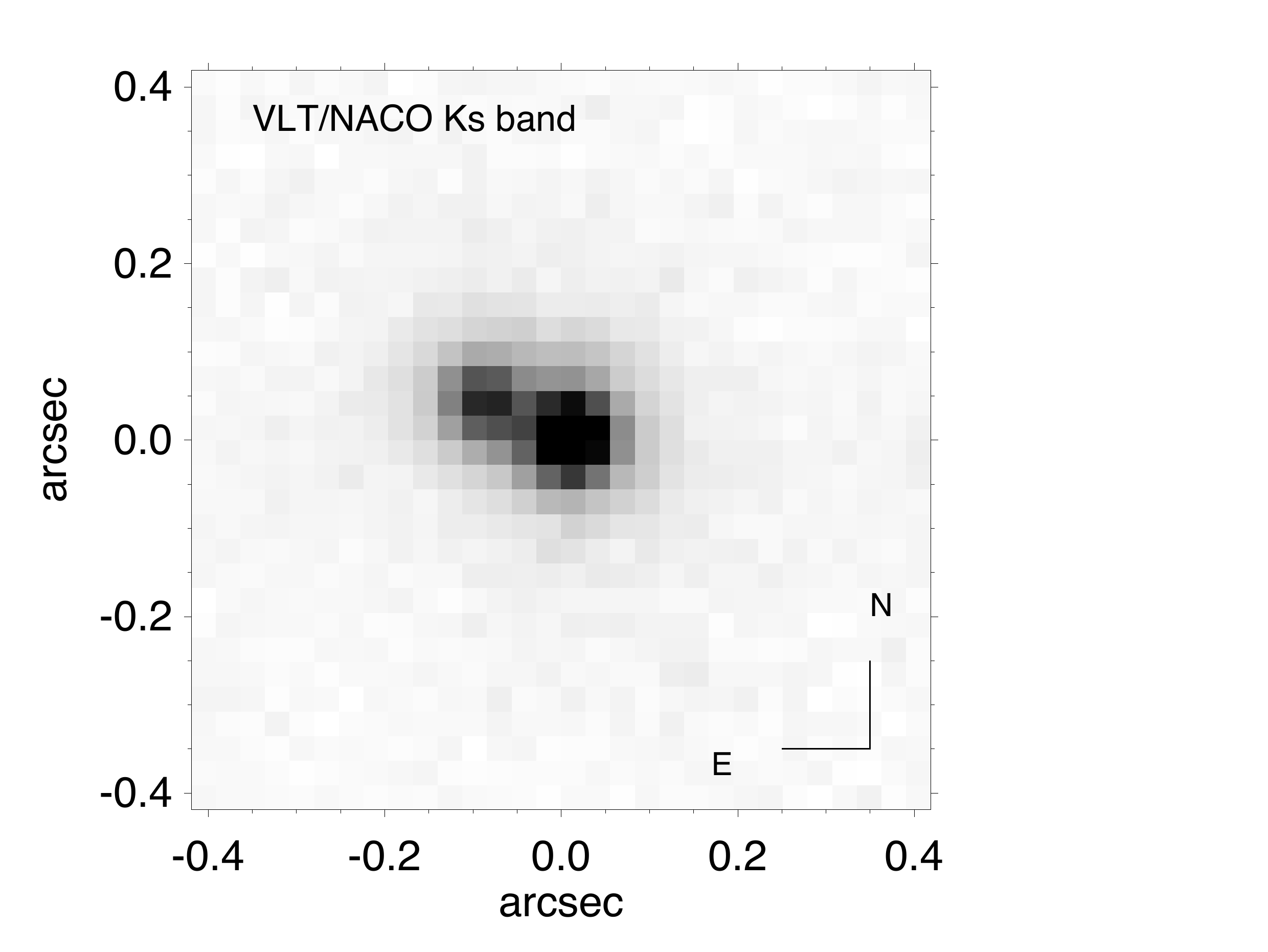}\vspace{-0.1cm}
        \hspace*{2.0cm}  19\,/\,06\,/\,2009
      \end{minipage} \hspace{0.15cm}
 \caption{\label{det} Timeline for the observations of SDSS J2052-1609\,AB. The orientation is the same for all images. The 2MASS image shows the unresolved system and the circle indicates the 4.8\arcsec \,region around the target in which no other object is visible. The object in the north-west corner of the image is the PSF reference object 2MASS\,J205234-160913. The HST/NICMOS and VLT/NACO images show the resolved system where SDSS J2052-1609\,B is the component north-east of SDSS J2052-1609\,A.}

\end{figure*}
%

As part of a high-resolution near-IR imaging survey, which had the aim to find tight T dwarf binaries, we observed \object{SDSS J205235.31-160929.8} (hereafter SDSS J2052-1609) and resolved it into a new close L/T transition binary system. Originally, SDSS J2052-1609 was identified as an apparent single object in a SDSS survey by \citet{chiu2006} and classified as a T1\,$\pm$\,1 dwarf. Furthermore, SDSS J2052-1609 has a measured proper motion of $\mu$ = 0.483\,$\pm$\,0.022 arcsec/yr (Faherty 2010, in preparation, see \citealp{burgasser2010}). 
Just recently \citet{burgasser2010} identified SDSS J2052-1609 as a binary candidate using a spectral template fitting technique. Subsequently it was ranked as a strong binary candidate, since the best-fit composite spectrum returned a much better fit than any single dwarf template. SDSS J2052-1609 was best fitted by a composite spectrum of the L7 dwarf \object{SDSS J115553.86+055957.5} (hereafter SDSS J1155-0559, \citealp{knapp2004}) and the T2 dwarf \object{2MASS J11220826-3512363} (hereafter 2MASS J1122-3512, \citealp{tinney2005}) and \citet{burgasser2010} estimated average component types of L7.5\,$\pm$\,0.6 and T2\,$\pm$\,0.2 for the SDSS J2052-1609\,AB system. 

Here we present the observations which resolved SDSS J2052-1609\,AB using NACO at the Very Large Telescope (VLT).

\section{Observation and data reduction}

\subsection{NACO}
The near-infrared imaging observations of the T1 dwarf SDSS J2052-1609 were obtained with 
NACO at the Very Large Telescope (VLT) on Cerro Paranal. The NACO system provides 
high-resolution AO assisted imaging in the near-infrared at an 8-meter class telescope. The 
observations were carried out in service mode as a part of our T dwarf high-resolution imaging campaign. SDSS J2052-1609 was observed on June 19, 2009 in \emph{H} (1.65 $\mu$m) and \emph{K$_\mathrm{S}$} (2.15 $\mu$m) broad-band filters using the CONICA S27 camera. The 
S27 camera provides a field of view (FoV) of 28\arcsec\,$\times$\,28\arcsec \,and a pixel scale of 0.0271\arcsec. For AO correction the wave front sensing was performed on a reference source chosen from the GSC-II (V2.2.01). The star S3313312149 has \emph{R}\,=\,12.85 mag and is 37.3$\arcsec$ away from SDSS J2052-1609. During the observation the target was at a mean airmass of 1.148 in \emph{H} and 1.104 in \emph{K$_\mathrm{S}$}, with average seeing conditions between 0.8\arcsec and 0.6\arcsec\,in \emph{H} and \emph{K$_\mathrm{S}$}, respectively. The observations were executed using a 8 point dither pattern, to allow for sky subtraction and bad pixel correction, resulting in total integration times of 1200\,s in \emph{H}\,-\,band and 960\,s in \emph{K$_\mathrm{S}$}\,-\,band. Standard data reduction steps as flat-fielding, sky subtraction and bad pixel corrections were performed before averaging the eight dither positions with the Eclipse \emph{jitter} \citep{Devil} software package. The final reduced NACO images of SDSS J2052-1609\,AB are shown in Figure\,\ref{det} and represent the detection images of the resolved system. 

For the determination of the relative photometry (flux ratio) and astrometry (separation and position angle (PA)) of the new system we used the IDL\,-\,based simultaneous PSF\,-\,fitting algorithm from \citet{bouy2003}, adapted to the VLT/NACO data. As PSF reference star we used the nearby point source 2MASS\,J205234-160913 which was observed in the same FoV as our target and thus under the same observing conditions. To determine the statistical error of this fit, the algorithm was also applied to each individually reduced image. Finally the results per filter were averaged and the uncertainties were calculated from the standard deviation. To account for the systematic errors an uncertainty of 1\% for the pixel scale and 0.4$^{\circ}$ for the orientation (e.g. \citealp{Koehler08, Eggenberger07}) was added in the end.

\begin{table*}[t!]
\caption{\label{phot} Photometric properties of the resolved SDSS2052-1609\,AB 
system}
\centering
\begin{tabular}{lccccc}
\noalign{\smallskip}
\hline\hline
\noalign{\smallskip}
Filter   & flux ratio & $\Delta$ mag & Comp. A & Comp. B & A\,+\,B \\
          &            & [mag]        & [mag]   & [mag]   & [mag] \\
\noalign{\smallskip}
\hline
\noalign{\smallskip}
F110W\,$^{\mathit{a}}$   & 0.787\,$\pm$\,0.012 & 0.26\,$\pm$\,0.02 & 17.77\,$\pm$\,0.03 & 18.02\,$\pm$\,0.03 & 17.14\,$\pm$\,0.02 \\
F170M\,$^{\mathit{a}}$   & 0.576\,$\pm$\,0.005 & 0.60\,$\pm$\,0.01 & 15.86\,$\pm$\,0.02 & 16.46\,$\pm$\,0.02 & 15.37\,$\pm$\,0.02\\[0.3ex]
\emph{J}\,$^{\mathit{b}}$ & {\bf 0.908\,$\pm$\,0.149} &{\bf 0.10\,$\pm$\,0.18} & {\bf 17.00\,$\pm$\,0.13} & {\bf 17.10}\,$\pm$\,{\bf 0.13} & {\bf 16.30\,$\pm$\,0.12} \\
\emph{H}\,$^{\mathit{c}}$ & 0.591\,$\pm$\,0.007 & 0.57\,$\pm$\,0.01 & 15.92\,$\pm$\,0.12 & 16.49\,$\pm$\,0.12 & 15.41\,$\pm$\,0.12 \\
\emph{K$_\mathrm{S}$}\,$^{\mathit{c}}$ & 0.513\,$\pm$\,0.005 & 0.72\,$\pm$\,0.01 & 15.57\,$\pm$\,0.15 & 16.30\,$\pm$\,0.15 & 15.12\,$\pm$\,0.15 \\
\noalign{\smallskip}
\hline
\noalign{\smallskip}
\emph{J\,--\,H} 	  & & & 1.08\,$\pm$\,0.05  & 0.61\,$\pm$\,0.05  & 0.88\,$\pm$\,0.04 \\
\emph{H\,--\,K$_\mathrm{S}$} & & & 0.35\,$\pm$\,0.19  & 0.19\,$\pm$\,0.19  & 0.29\,$\pm$\,0.19 \\
\emph{J\,--\,K$_\mathrm{S}$} & & & 1.43\,$\pm$\,0.20  & 0.81\,$\pm$\,0.20  & 1.17\,$\pm$\,0.19 \\
\noalign{\smallskip}
\hline 
\end{tabular}
\begin{list}{}{} 
\item[$^{\mathit{a}}$] the NICMOS photometry is on the Arizona Vega System ($M_{Vega}$\,=\,0.02).
\item[$^{\mathit{b}}$] due to the lack of direct observations, the \emph{J}-\,band magnitudes are based on the NICMOS photometry which were transformed from the Vega system into the 2MASS system as described in the text.
\item[$^{\mathit{c}}$] based on 2MASS \emph{H, K$_\mathrm{S}$} photometry for the unresolved sources.
\end{list}
\end{table*}

\subsection{HST and 2MASS}

Having resolved SDSS J2052-1609 as a candidate binary we checked the 2MASS and HST data archives for previous epoch data. We downloaded the available 2MASS \emph{J}, \emph{H} and \emph{K$_\mathrm{S}$} imaging data which was taken on June 01,\,1999. The Two Micron 
All-Sky Survey provides near-infrared imaging data with a pixel scale of 1\arcsec. Since SDSS J2052-1609 has a measured proper motion of 0.483 arcsec/yr, the dwarf should have moved by approximately 4.8\arcsec \,during the last 10 years, corresponding to $\sim$\,4.8 pixel in the 2MASS image. The 2MASS images show no object around SDSS J2052-1609 within a radius of $\sim$ 5\arcsec. 

The HST archive proved to be more fruitful. NICMOS imaging observation of SDSS J2052-1609\,AB have been obtained under program GO 11136 (PI: M.Liu). The observation were carried out on June 24, 2008 and used the NICMOS1 (NIC1) camera with a FoV of 11\arcsec\,$\times$\,11\arcsec and a pixel scale of 0.043\arcsec. Observations of SDSS J2052-1609\,AB are available in several filters. Here we used the \emph{F110W} and \emph{F170M} filters, with total integration times of 12\,s and 60\,s, respectively. The telescope orientation angle during all observations was -27.9068$^{\circ}$ (E of N). 

The data analysis is based on the pipeline-reduced frames provided by the HST archive. To derive the magnitudes from the HST/NIC1 images we applied aperture photometry to measure the total flux of the system, since the two components of the binary were too tight to derive their absolute photometry separately. With IDL-based routines the count rate within an aperture of 11.5 pixel was measured, corrected to a nominal infinite aperture and converted into the ``Arizona'' Vega photometric system (Table\,\ref{phot}), using the photometric keywords and flux zero points provided on the STScI webpage\footnote{http://www.stsci.edu/hst/nicmos/performance/photometry/\\postncs\_keywords.html}.

Similar to the VLT/NACO data we applied the PSF\,-\,fitting algorithm from \citet{bouy2003} 
to derive the astrometric parameters of the binary as well as the flux ratio of the two 
components. This time we built and used a library of 6 reference PSFs: one TinyTim synthetic 
PSF and five natural PSFs from other HST/NICMOS programs that targeted brown dwarfs of similar spectral type (L7 - T1.5) in the corresponding filters (GO 10143 and 10879, PI: I.Reid). 
For a precise measurement of the separation, PA and flux ratio the algorithm was used on each of the 4  images per filter. The final results are the average per filter of the individual results and the error was calculated from the standard deviation. 

\section{Results}
\subsection{Photometry}
Table\,\ref{phot} lists the flux ratio and the corresponding magnitude difference obtained from the binary fitting procedure. For the \emph{H} and \emph{K$_\mathrm{S}$} filter the 2MASS magnitudes of the combined system are given in column `A\,+\,B', while for the NICMOS filters the aperture photometry result is listed. Individual magnitudes for the two components were then calculated from these system magnitudes using the flux ratio determined during the binary fitting procedure. In order to estimate the individual \emph{J}\,-\,band magnitudes of the components in the 2MASS system, we applied the relation defined by \citet{burgasser2006c}. Thereby, the NICMOS \emph{F110W} and \emph{F170M} magnitudes are transformed into MKO \emph{J}\,- and \emph{H}\,-\,band magnitudes. Afterwards the MKO \emph{J}\,-\,\emph{H} colors were transformed into the 2MASS system\protect\footnote{http://www.astro.caltech.edu/$\sim$jmc/2mass/v3/transformations/} to calculate the 2MASS \emph{J}\,-\,band magnitudes (given in bold in Table \ref{phot}). For comparison, the derived \emph{J}\,-\,band magnitude of the unresolved system (\emph{J}\,=\,16.30\,$\pm$\,0.12\,mag) is within its errors in very good agreement to the published 2MASS magnitude (\emph{J}\,=\,16.33\,$\pm$\,0.12\,mag). Finally, the given\, \emph{J}\,-\,band flux ratio was derived from the separately transformed component magnitudes.
\begin{figure}
\centering 
\includegraphics[scale=0.3]{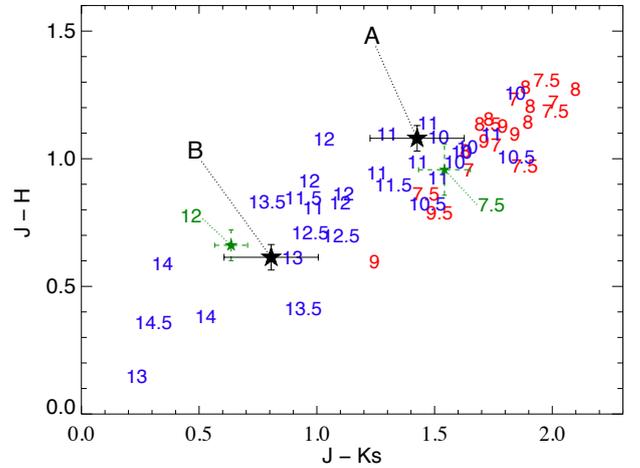}
 \caption{Location of SDSS J2052-1609\,A and B in the \emph{J\,-\,K$_\mathrm{S}$} vs. 
\emph{J\,-\,H} plane relative to known dwarfs from the \emph{DwarfArchives}. The numbers indicate the individual spectral types from 7\,=\,L7 to 14.5\,=\,T4.5. The two green numbers indicate the location of SDSS\,J1155+0559 (L7.5) and SDSS\,J1122-3512 (T2), the two best-fit spectral composite components derived by \citet{burgasser2010}.}
 \label{fig:jkjh} 
\end{figure}

\subsection{Spectral type and distance}

The derived colors of the binary components are given in Table\,\ref{phot}. A comparison to  mean \emph{JHK$_\mathrm{S}$} colors of L and T dwarfs, calculated from the dwarfs listed at the \emph{Dwarf Archives}\footnote{http://www.DwarfArchives.org} (provided in the 2MASS photometric system) shows a good agreement with spectral types of T0\,-\,T1 for component A and spectral types of T2\,-\,T3 for the B component (see also Figure~\ref{fig:jkjh} for illustration). An additional comparison with the average \emph{J\,-\,K$_\mathrm{S}$} colors vs. SpT in Table 5 of \citet{Faherty09} yields the same result. In fact, the colors of SDSS J2052-1609\,A are almost identical to the colors of \object{SDSSp J083717.22-000018.3} (hereafter SDSS J0837-0000:  \emph{J\,-\,H}\,=\,1.11\,$\pm$\,0.28\,mag, \emph{H\,-\,K$_\mathrm{S}$}\,=\,0.31\,$\pm$\,0.23\,mag, \emph{J\,-\,K$_\mathrm{S}$}\,=\,1.43\,$\pm$\,0.26\,mag), which has an assigned optical SpT of T0 \citep{kirkpatrick2008} and T1 in the near-IR \citep{burgasser2006b}. A different approach to estimate the spectral types has been done using the SpT\,-\,NICMOS color relation given in \citet{burgasser2006c}. For the B component the calculation yields a spectral type of T2, while for component A the spectral type is roughly T1. Thus, these estimated spectral types are in good agreement with the spectral types yielded by the \emph{JHK$_\mathrm{S}$} color comparison. 
%
\begin{figure}
\centering 
\includegraphics[scale=0.32,angle=90]{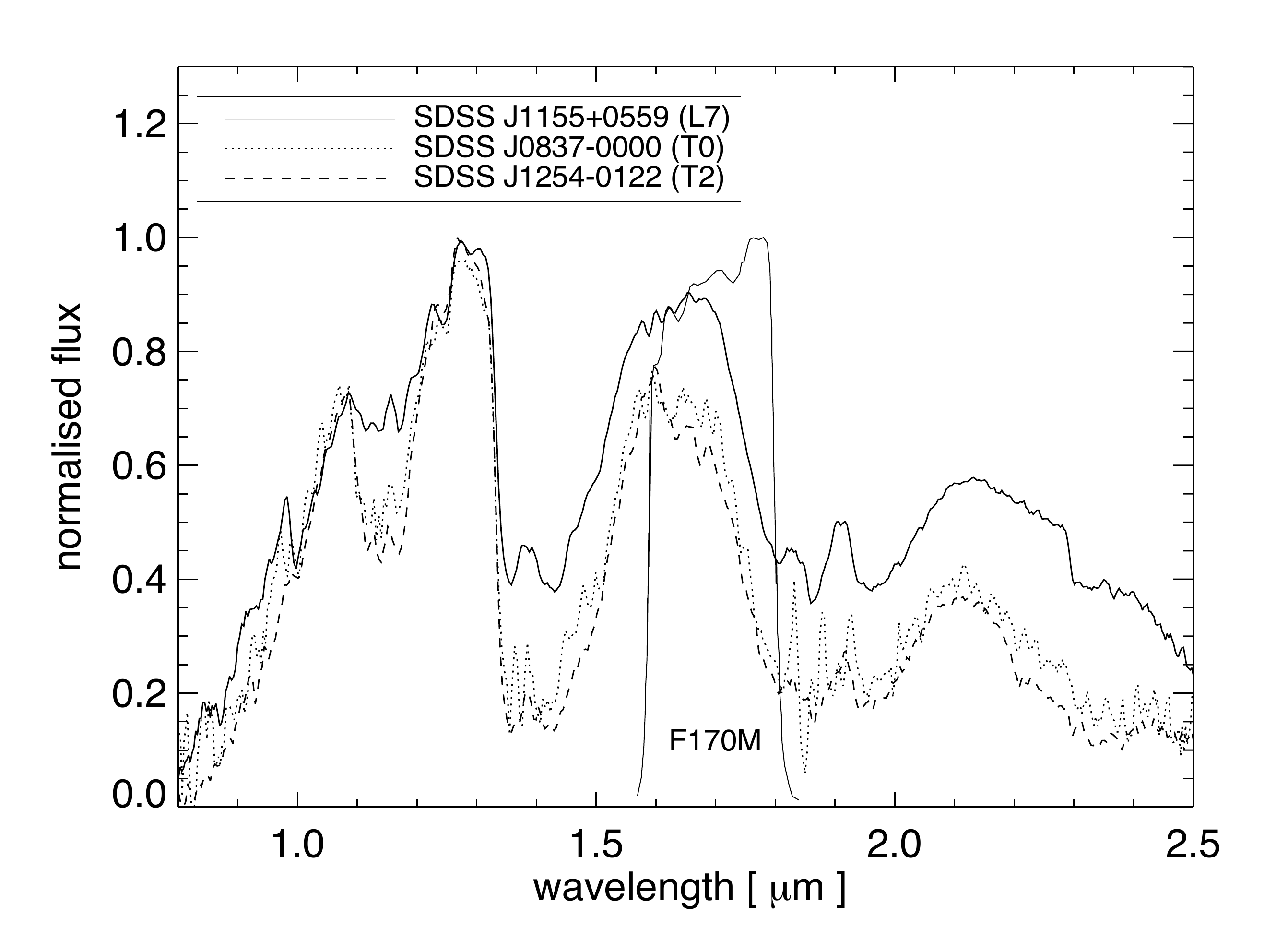}
  \caption{Comparison of the flux distribution in the NICMOS/NIC1 \emph{F170M} filter bandpass between a blue L7, a T0 and the T2 spectral standard. All spectra are normalized to the peak flux in the 1.0\,-\,1.3 $\mu$m range.}
 \label{fig:f170m}
\end{figure}

In contrast, the spectral binary fitting of the unresolved SDSS J2052-1609\,AB by \citet{burgasser2010} returned spectral types of L7.5\,$\pm$\,0.6 and T2\,$\pm$\,0.2. While the spectral type of the B component agrees with our estimation, the spectral type of the A component differs by 3\,-\,4 subclasses. The A component of their best-fit composite template, SDSS J1155+0559, is an L7 dwarf with unusually blue near-IR colors (\emph{J\,-\,H}\,=\,0.96\,$\pm$\,0.11\,mag, \emph{H\,-\,K$_\mathrm{S}$}\,=\,0.59\,$\pm$\,0.10\,mag, \emph{J\,-\,K$_\mathrm{S}$}\,=\,1.54\,$\pm$\,0.11\,mag). Figure~\ref{fig:f170m} compares the spectra of SDSS J1155+0559 (L7, \citealp{burgasser2010}), SDSS J0837-0000 (T0, \citealp{burgasser2006b}) and the T2 spectral standard \object{SDSS J125453.90-012247.4} (SDSS J1254+0122, \citealp{burgasser2004}), which were derived from the SpeX prism spectral libraries\footnote{http://www.browndwarfs.org/spexprism}. It illustrates their flux emitted in the \emph{F170M} filter which is centered at the methane absorption band. The assumption of SDSS J2052-1609\,A being an unusually blue late type L dwarf with no methane absorption could explain the large difference in the \emph{F170M} magnitudes of component A and B, which is not expected if the spectral types are as similar as T1 and T2. With only the near-IR colors to rely on, we assign preliminary spectral types of T1\,$^{+\,1}_{-\,4}$ and T2.5\,$\pm$\,1 for SDSS J2052-1609\,A and B, respectively, in which the larger uncertainty of the A component accounts for the yet unexplained large $\Delta$\,F170M mag of the components. Only resolved spectroscopy of the system components can unambiguously determine their spectral types.

Using the approximate spectral types we can estimated the distance by employing the 
relation between absolute magnitude and spectral type from \citet{looper2008}. While the 
unresolved system had a photometric distance of 22\,$\pm$\,1.7\,pc, the resolved system will 
be further away. For each of the individual components we calculated the distance based on 
the \emph{H} and \emph{K$_{S}$} magnitudes (see Table\,\ref{dist}) and the uncertainties were derived from the photometric, as well as the absolute magnitude vs.\, SpT relation uncertainties. Assuming spectral types T1 and T2 for SDSS J2052-1609\,A and B we derive an average distance of 31.8\,$\pm$\,2.5\,pc for the binary system. Given the large uncertainty in the spectral type of component A the average distance may increase up to 35.8\,$\pm$\,3.4\,pc if we assume a spectral type of L7.5.
\begin{table}[t!]
\centering
\caption{\label{dist} Estimated distances of SDSS J2052-1609\,AB in parsec. }
\centering
\begin{tabular}{ccccc}
\hline\hline
\noalign{\medskip}
    &  A (L7.5)   & A (T1)  & B (T2) & A\,+\,B (T1) \\
\noalign{\smallskip}
\hline
\noalign{\smallskip}
\emph{J} & & & & 22.3\,$\pm$\,3.2 \\
\emph{H} & 31.8\,$\pm$\,6.4 & 27.2\,$\pm$\,4.0 & 34.6\,$\pm$\,5.0  &
21.6\,$\pm$\,3.1 \\
\emph{K$_\mathrm{S}$} & 37.2\,$\pm$\,8.9 &28.6\,$\pm$\,4.8 & 36.7\,$\pm$\,6.1 &
23.2\,$\pm$\,3.7 \\
\noalign{\smallskip}
\hline
\end{tabular}
\end{table}

\subsection{System parameters}

To derive the most reliable astrometric parameters from the 2008 HST/NICMOS and 2009 
VLT/NACO observations, we only used the \emph{F110W} and \emph{K$_\mathrm{S}$}\,-\,band 
data to determine the first orbital parameters, since those two filters provide results 
which are notably better than the diffraction limit in these filters. In addition, the NACO 
observations in the \emph{K$_\mathrm{S}$}\,-\,band are obtained at a much better Strehl 
ratio (with an average value of 69\,\% compared to 54\,\% in \emph{H}\,-\,band).
The derived separations and position angles are listed in Table~\ref{sep}. The observations 
are separated by a little bit less than one year. During this time the separation between 
the two components changed from 102.7\,mas to 100.9\,mas. With this total decrease of only 
1.8\,$\pm$\,1.5\,mas the separation between components A and B remained almost constant 
over the period of one year. This leads to the conclusion that the two components must be 
co-moving, since SDSS J2052-1609\,AB has a reported proper motion of 
$\mu$\,=\,0.483\,$\pm$\,0.022$\arcsec$/yr. In the same time interval the position angle 
changed by 16.9\degr$\pm$\,0.5\degr. \\ 
\begin{table}[b!]
\centering
\caption{\label{sep} Astrometric parameters for the SDSS J2052-1609\,AB system.}
\centering
\begin{tabular}{l l c c }     
\noalign{\smallskip}
\hline\hline
\noalign{\medskip}
& & HST/NIC1 &VLT/NACO\\
Parameter & & \emph{F110W}	 & \emph{K$_\mathrm{S}$}\\
& & (24\,/\,06\,/\,2008)\,$^{\mathit{a}}$ & (19\,/\,06\,/\,2009)\\
\noalign{\medskip}
\hline
\noalign{\smallskip}
\noalign{\smallskip}
Separation & $\rho$ [mas] &  102.7 $\pm$ 1.1 & 100.9 $\pm$ 1.0 \\[0.5ex]
Position Angle & $\theta$ [deg]   &  50.4 $\pm$ 0.2 &  67.3 $\pm$ 0.5 \\[1ex]
estimated distance & $d$ [pc] &  \multicolumn{2}{c}{31.8 $\pm$ 2.3}\\ [0.5ex]
Semi-major axis & $a$ [AU]   & \multicolumn{2}{c}{\hspace{-2.5ex}$\ge$\, 3.2 $\pm$ 0.5}\\ [0.5ex]
Orbital period & $P$ [years]   & \multicolumn{2}{c}{21 $\pm$ 1}\\[0.5ex]
System mass & $M_{tot}$ [$M_\mathrm{Jup}$]  & \multicolumn{2}{c}{\hspace{-1.5ex}$\ge$\,  78}\\ 
\noalign{\smallskip}
\hline
\end{tabular}
\begin{list}{}{} 
\item[$^{\mathit{a}}$] \emph{HST} observation by M.Liu that are present in the public archive.
\end{list}

\end{table}
The non-significant change in the separation allowed for a first assumption of a circular 
orbit seen face-on and to estimate the orbital period of the system. Thereby the orbital 
movement of 16.9$^{\circ}$ per year translates into an orbital period of 21\,$\pm$1\,yr.
With an averaged separation of 101.8\,$\pm$\,1.5 mas and the new determined photometric distance of 31.8\,pc the semi-major axis of the system corresponds to a projected 
separation of 3.2\,$\pm$\,0.5\,AU. Using Kepler's law this yields a first estimate of the 
total system mass of $\approx$\, 78\,M$_{Jup}$. 
Depending on the inclination (if {\it i}\,$>$\,0) and eccentricity ({\it e}) of the system, the true semi-major axis is on average larger than the observed separation \citep{fischer1992}, resulting in an increase of the total system mass. Only in the case of an eccentric orbit seen face-on and observed close to the periastron the estimated total system mass would be an upper limit. An on-going astrometric monitoring program will help to better determine these orbital parameters in the future. 

\section{Summary}

VLT/NACO observations of SDSS J2052-1609 resolved the brown dwarf into a close binary system with a separation of only $\sim$\,101.8\,mas. Previous, unpublished observations conducted with HST/NICMOS confirm the binary nature of the system and allow for an first estimation of the total system mass. Assuming a circular orbit with an orbital period of 21\,$\pm$\,1 years we estimate a system mass of $\geq$\,78\,M$_{Jup}$. The near-IR colors of the individual components suggest spectral types of T1\,$^{+\,1}_{-\,4}$ and T2.5\,$\pm$\,1, respectively. 

A divergent estimate of the spectral types comes from spectral binary fitting \citep{burgasser2010}, suggesting spectral types of L7.5 and T2 for component A and B. Their best-fit composite template consists of two late type with unusually blue near-IR colors. Our obtained photometry  is in agreement with both scenarios: a `normal' T\,+\,T dwarf binary or a L/T transition system consisting of a blue L and an early T dwarf. 
 
Upcoming resolved spectroscopy with SINFONI at the VLT will finally determine the 
spectral types of the system.

\begin{acknowledgements}
M.B.Stumpf and W. Brandner acknowledge support by the \emph{DLR Verbundforschung} project numbers 50 OR 0401 and 50 OR 0902. We would like to thank the anonymous referee for the constructive comments, that helped to improve the paper.
This publication makes use of data products from the Two Micron All Sky Survey, which is a joint project of the University of Massachusetts and the Infrared Processing and Analysis Center/California Institute of Technology, funded by the National Aeronautics and Space Administration and the National Science Foundation. This research has benefitted from the M, L, and T dwarf compendium housed at DwarfArchives.org and maintained by Chris Gelino, Davy Kirkpatrick, and Adam Burgasser. This research has benefitted from the SpeX Prism Spectral Libraries, maintained by Adam Burgasser at http://www.browndwarfs.org/spexprism.
\end{acknowledgements}

\bibliographystyle{aa} 
\bibliography{sdss} 
\end{document}